\documentclass{article}
\usepackage{spconf,amsmath,amssymb,graphicx,booktabs,array,tabularx,multirow,hyperref,cuted,capt-of}
\newcommand{\pos}[1]{\left[#1\right]_{+}}

\newcommand{\Lrel}{\mathcal{L}_{\mathrm{rel}}}
\newcommand{\Ldam}{\mathcal{L}_{\mathrm{dam}}}

\newcommand{\kS}{\kappa_S}
\newcommand{\kR}{\kappa_R}

\title{RELATIONAL SYNTHESIS: STRUCTURE-MEDIATED CONCATENATIVE SYNTHESIS FOR FOLEY AND RETRIEVAL-AUGMENTED AUDIO GENERATION}
\name{Keren Shao, Ayaka Kawano, Shlomo Dubnov}
\address{University of California San Diego, La Jolla, CA, USA}

\begin{document}
\flushbottom
\ninept
\maketitle

\begin{abstract}
We ask: given a retrieved source audio $S$ and a separate reference audio $R$, can we synthesize novel audio $Y$ out of this pair $(S,R)$ such that $Y$ remains acoustically consistent with $S$, while not persistently copying segments of $S$ or $R$? The first clause is a well-known goal in Foley audio production, and the second is a well-known issue in neural RAG when $S$ and $R$ are naively injected into neural generators. We show that both clauses can be addressed simultaneously using a method we coin \textbf{relational synthesis}, a variation of concatenative synthesis where target cost is replaced by a relational Gromov-like structural cost. Rather than imitating the content of $R$, relational synthesis exploits it from the ``other side of the hill'': it transfers the temporal structure and directed amplitude motion of $R$ to reorganize and concatenate the grains of $S$ in a novel manner that protects $S$'s acoustic information. Our experiments show that relational synthesis integrates naturally with neural RAG and produces Foley audio that performs well on metrics measuring temporal agreement, acoustic fidelity, and leakage persistence, while maintaining distribution-level quality and text alignment.
\end{abstract}

\begin{keywords}
relational synthesis, concatenative synthesis, Foley sound synthesis, retrieval-augmented generation, temporal control
\end{keywords}

\section{Introduction}
Foley synthesis must satisfy two coupled demands: \emph{what the sound is}, for example event class, material, timbre, acoustic identity, and \emph{when and how it happens}, for example onset/offset, rhythm, intensity, temporal structure. Temporal-event and vocal-imitation control make this split explicit \cite{chung2024tfoley,okamoto2024vocal}. Retrieval-augmented audio generation (RAG) supplies complementary acoustic evidence that may be difficult to specify in text \cite{yuan2024reaudioldm,yang2025audiobox,fang2026acfoley}, but direct audio conditioning introduces a different risk: useful retrieved/reference material may persist as copied audio rather than merely guide generation \cite{bralios2024replication}.

Let $S$ be the retrieved source, $R$ an independent temporal reference, and $Y$ the output. We seek
\[
\mathcal A(Y)\approx\mathcal A(S),\qquad \mathcal T(Y)\approx\mathcal T(R),\qquad \Pi(Y,S),\Pi(Y,R)\ \text{small}.
\]
Here $\mathcal A$ denotes event-class acoustic identity, $\mathcal T$ temporal organization, and $\Pi$ persistent local replication. Thus $S$ should supply what the event sounds like, $R$ should supply how it unfolds, and neither waveform should simply survive inside $Y$.

Concatenative synthesis is attractive because it recombines source material rather than parametrically injecting $S$ or $R$ directly \cite{schwarz2007corpus}, making the $\Pi$ constraints comparatively easier to satisfy. The obvious candidate is pointwise target matching, but this creates a granularity dilemma. Fine grains retain enough degrees of freedom to follow $\mathcal T(R)$, yet force local slices from potentially different acoustic spaces, for example environmental sound and vocal imitation, to match directly; differences in local dynamics and variance can make onset/offset and joins awkward, while larger reference structure is ignored. Longer segments preserve more source-native continuity and therefore favor $\mathcal A(S)$, but remain local matches, sacrifice framewise temporal freedom as segments lengthen, and let longer contiguous spans of $S$ survive, working against $\Pi$.

We therefore change \emph{how grains are matched}. \textbf{Relational synthesis}\footnote{Audio examples are available at \url{https://smoothken.github.io/relational_synth_demo/}.} keeps framewise source selection, but replaces reference-to-source pointwise matching $d(r_t,s_{x_t})$ with constraints on within-domain relations. Crucially, this makes the $\mathcal A$-$\mathcal T$ trade-off less rigid: recurrence structure from $R$ specifies non-local similarity and dissimilarity patterns, while relational amplitude structure of $R$ transfers the relative strength and direction of rises and falls without requiring source and reference envelopes to match in absolute value, leaving room for the solver to retain acoustic fidelity when selecting grains of $S$. The same framewise search can therefore pursue $\mathcal A(Y)\approx\mathcal A(S)$ and $\mathcal T(Y)\approx\mathcal T(R)$ together while recombination continues to limit persistent copy-through, $\Pi(Y,S)$ and $\Pi(Y,R)$. To solidly realize this synthesis principle, we formulate weighted recurrence structures with three models: contextual recurrence graph (CRG), factor oracle (FO), and variable-Markov oracle (VMO). We then specify the relational amplitude structure with directed amplitude motion (DAM), and solve the resulting grain-selection optimization with an annealed discrete solver.

\begin{figure*}[t]
\centering
\includegraphics[width=0.8\textwidth]
{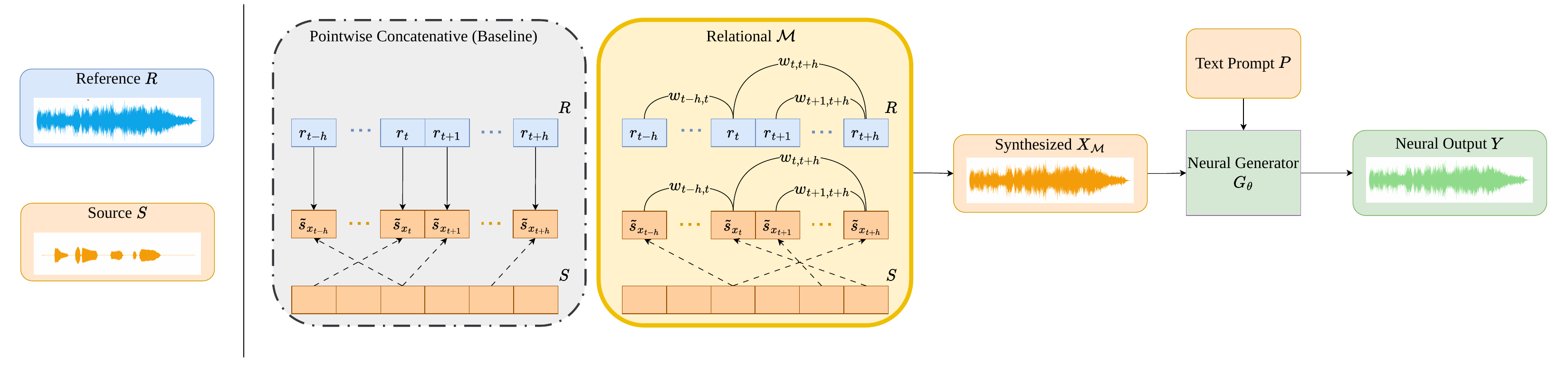}
\vspace{-2mm}
\caption{Relational synthesis pipeline. Reference $R$ specifies temporal relations while source $S$ supplies selectable acoustic material. Method $\mathcal M$ reorganizes source grains into carrier $X_{\mathcal M}$, which conditions pretrained generator $G_\theta$ together with prompt $P$ to produce neural output $Y$. In our experiments, $R$ is a vocal imitation paired with target $T$ of the same class as $S$.}
\label{fig:pipeline}
\vspace{-2mm}
\end{figure*}

\section{Methods}
All variants implement the same coupling principle: a method $\mathcal M$ defines relational pressures of $R$ and a framewise path through $S$ realizes them acoustically. Let $S=(s_1,\ldots,s_M)$ and $R=(r_1,\ldots,r_N)$ be 180-ms source/reference grains at 100-ms hop. The synthesis variable $x_t\in\{1,\ldots,M\}$ selects the source grain realizing reference time $t$, and $X_{\mathcal M} =\operatorname{OLA}(s_{x_1},\ldots,s_{x_N})$. For any grain $g$, the shared heuristic acoustic embedding is
\[
z=\phi(g)=\operatorname{norm}[\overline{\mathrm{MFCC}},\overline{\Delta\mathrm{MFCC}},\mathrm{Env}_8,\log RMS],
\]
with pooled MFCC/delta trajectories, a coarse normalized envelope, and log RMS, followed by standardization and $\ell_2$ normalization. This embedding consists of handcrafted acoustic features commonly used in acoustic modeling and can be readily replaced by, e.g., modern self-supervised acoustic embeddings. We leave this exploration to future work. Let $\kS(i,j)\in[0,1]$ denote the cosine similarity between the embeddings $z(s_i)$ and $z(s_j)$. Each reference structure produces positive and negative edges $E^+,E^-$ and weights $w^+,w^-$. All three recurrence structure specifiers, CRG, FO, and VMO, share
\begin{equation}
\begin{aligned}
\Lrel(x)={}&\sum_{(t,u)\in E^+}w^+_{tu}(1-\kS(x_t,x_u))^2\\
&+\lambda_-\sum_{(t,u)\in E^-}w^-_{tu}\pos{\kS(x_t,x_u)-m}^{2}.
\end{aligned}
\label{eq:rel}
\end{equation}
Here $\lambda_-\ge0$ weights negative relations and $m\in[0,1]$ is their similarity margin. We use $\lambda_-=.35$ and $m=.55$. Thus $R$ determines \emph{which times should relate}, while the solver uses source self-similarity to realize those relations without a reference-source pointwise acoustic distance. This construction is Gromov-like in that incomparable observations are coupled through intra-domain relations rather than a pointwise ground metric \cite{cohen2021gdtw}. To keep the present scope focused, we restrict the formulation to reference-side relational structure rather than jointly modeling and matching relational structures in both $S$ and $R$, as in full-fledged Gromov-Wasserstein.

\subsection{Contextual Recurrence Graph}
The most direct relational structure is a contextual recurrence graph (CRG): repeated trajectories appear as off-diagonal paths in audio self-similarity matrices, motivating recurrence scores based on local diagonal context rather than isolated frame pairs \cite{mueller2006similarity, foote2001selfsim}. Let $\kR(t,u)$ be cosine similarities among the reference embeddings. With contextual radius $r$ and diagonal decay scale $\sigma$, the contextual recurrence significance $\Gamma_R(t,u)$ is computed as a weighted average of similarities along the local diagonal.

\begin{equation}
\Gamma_R(t,u)=\frac{\sum_{\delta=-r}^{r}e^{-|\delta|/\sigma}\kR(t+\delta,u+\delta)}{\sum_{\delta=-r}^{r}e^{-|\delta|/\sigma}}.
\end{equation}
After excluding $|t-u|\le r$, $E_t^+=\operatorname{TopK}_u\Gamma_R(t,u)$ and $E_t^-=\operatorname{BottomK}_u\Gamma_R(t,u)$, with $w^+_{tu}\propto\Gamma_R(t,u)^p$ and $w^-_{tu}\propto(1-\Gamma_R(t,u))^p$, where $p$ controls concentration on stronger edges. We set $(r,\sigma,K,p)=(4,2,8,2)$ in experiments. Diagonal averaging asks not merely whether $t$ and $u$ are similar, but whether their surrounding contexts evolve similarly.

\subsection{Factor-Oracle Graph}
Factor oracles provide a sparse variable-length view of recurrence: the suffix link of timestep \(t\) points to a timestep recognizing the longest repeated suffix (LRS) ending at \(t\) \cite{dubnov2007oracle,cont2007guidage}. Intuitively, the length of such repeated context provides a natural measure of relational evidence.

Because FOs operate on symbolic sequences while our audio embeddings are continuous-valued, we quantize the reference embeddings using greedy exemplar clustering,

\[
\text{symbol}_t=\begin{cases}
\arg\max_j\kR(t,e_j),&\max_j\kR(t,e_j)\ge\theta_{\mathrm{FO}},\\
\text{new symbol},&\text{otherwise}.
\end{cases}
\]
where $e_j$ denotes the current exemplar time of cluster $j$. We use $\theta_{\mathrm{FO}}=0.8$.

Once the clusters are formed, we construct the factor oracle on the symbolic sequence. Let $\operatorname{sfx}(t)$ denote the suffix-linked timestep and $\ell_t$ the LRS length at $t$. When $\ell_t\ge\ell_{\min}=3$, FO specifies the sparse positive relation $E_t^+=\{\operatorname{sfx}(t)\}$ with
\begin{equation}
w_t^+=\left[\min\left(1,\frac{\ell_t}{\ell_{0.9q}}\right)\right]^p,
\end{equation}
where $\ell_{0.9q}$ is the $90$th percentile of valid LRS lengths and robustly saturates very long repeats. Negative relations retain the contextual-recurrence construction of CRG. FO therefore concentrates positive relational pressure where the symbolic sequence provides evidence of non-trivial repeated context.

\begin{figure}[!t]
    \centering
    \includegraphics[
        width=0.45\textwidth,
        height=70mm,
        keepaspectratio
    ]{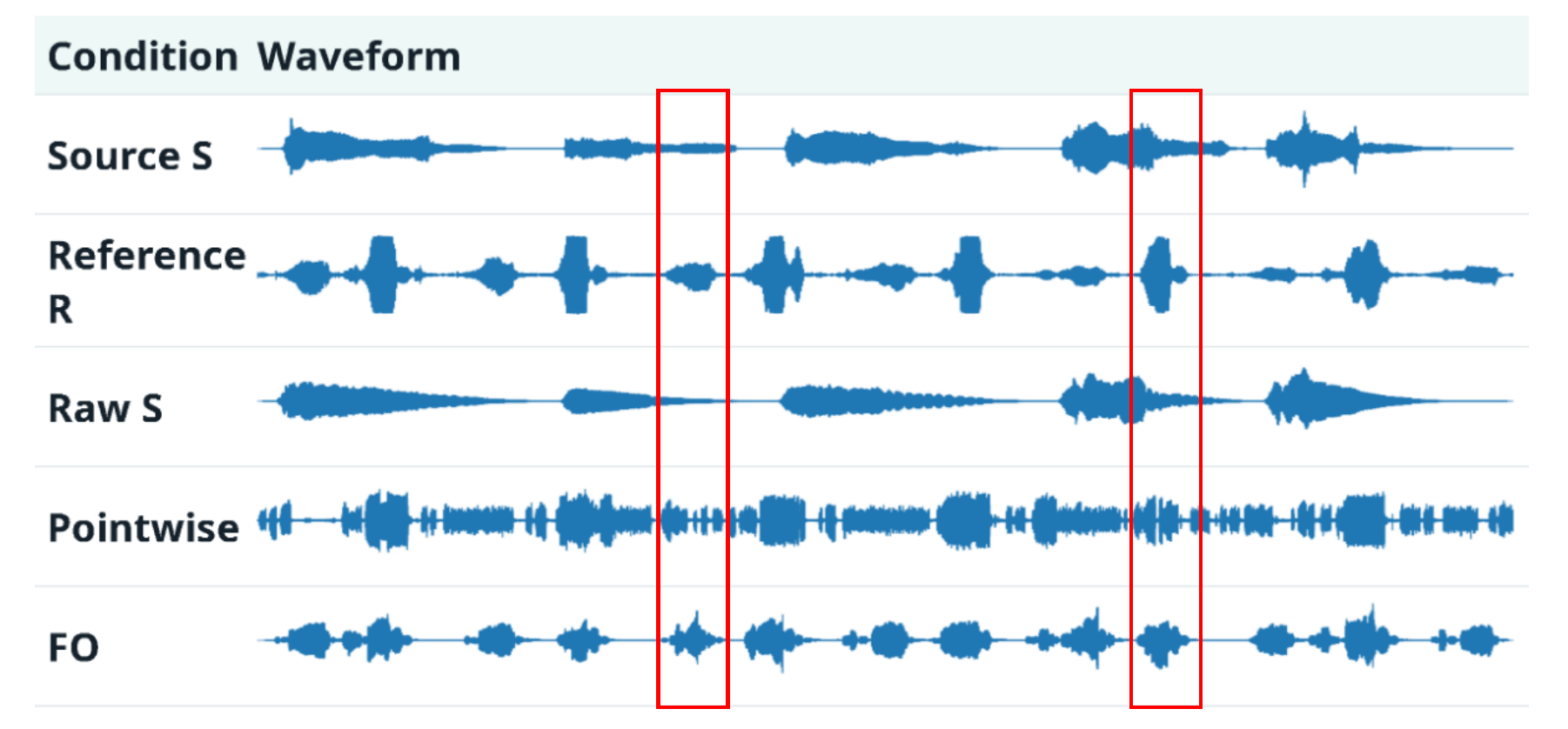}
    \vspace{-2mm}
    \caption{Out-of-dataset example with shakuhachi source $S$ and humpback-whale reference $R$. Raw source conditioning exhibits persistent source copying; pointwise concatenation reduces copying but poorly follows $R$'s temporal envelope. FO relational synthesis instead transfers the reference pattern while reorganizing source grains. Red boxes mark corresponding envelope structure.}

    \label{fig:ood}
\end{figure}

\subsection{Variable-Markov-Oracle Graph}
FO separates acoustic quantization from oracle construction. VMO instead determines acoustic clusters during oracle construction, jointly coupling symbolization with repeated-context discovery \cite{wang2014vmo,wang2015vmo}. Its clustering threshold $\theta$, rather than being fixed beforehand as in FO, is selected via information rate optimization $\theta^\star=\arg\max_\theta IR(Q_\theta)$, where $IR$ is information rate and $Q_\theta$ is the resulting VMO symbol sequence at threshold $\theta$. In practice, this is usually done via grid search $\theta \in [0.05, 1]$ with $0.05$ increments and selecting the run with the best $IR$, rather than in a continuously differentiable manner with respect to $\theta$.

To compare FO and VMO consistently, and in particular to isolate the effect of separate greedy clustering versus oracle-integrated clustering, we use the same definitions of $E_t^+$, $w_t^+$, $E_t^-$, and $w_t^-$ as in FO. In other words, CRG establishes relations through fixed-length context; FO introduces variable-length context through the longest repeated suffix; and VMO further enriches variable-order context with clusters optimized during oracle construction, at the cost of greater algorithmic complexity. These are parallel relational models rather than successive refinements: VMO is not assumed to dominate FO or CRG.

\subsection{Directed Amplitude Motion}

A similarity graph can have a hard time distinguishing opposite amplitude motions, such as natural onset- and offset-like trajectories,
\[
\text{quiet}\to\text{loud} 
\qquad\text{from}\qquad
\text{loud}\to\text{quiet}.
\]
across time-scales. To complement these recurrence relations, we therefore exploit another prominent relational structure of the waveform: directed amplitude motion. Let the log-RMS coordinates be
\[
a_t^R=\log RMS(r_t),\qquad a_i^S=\log RMS(s_i).
\]
In the experiments, $H=\{1,4,16,32\}$. For each lag $h\in H$, define the log-amplitude change

\vspace{-0.2mm}
$$
\Delta_h(t)=a_{t+h}-a_t,
$$
\vspace{-0.2mm}

and let $\widehat F_h$ denote the empirical CDF of such changes. Its quantile coordinate is

\vspace{-0.2mm}
$$
q_h(t)=\widehat F_h(\Delta_h(t)).
$$
\vspace{-0.2mm}

Intuitively, $q_h^R(t)\approx1$ indicates an unusually strong rise in the reference at time $t$, $q_h^R(t)\approx0$ an unusually strong fall, and $q_h^R(t)\approx\tfrac12$ a typical motion.

Consequently, for the synthesis to inherit the relative amplitude motion of $R$, we want the motion between $x_{t+h}$ and $x_t$, evaluated in the source as

$$
q_h^S(i,j)=\widehat F_h^S(a_j^S-a_i^S),
$$

to closely follow the \emph{relative extremeness} $q_h^R(t)$ for all $h\in H$ and $t$. In other words, we want
\vspace{-0.2mm}
\begin{equation}
\Ldam=\frac{1}{|H|}\sum_{h\in H}\sum_t\pos{\left|q_h^S(x_t,x_{t+h})-q_h^R(t)\right|-\varepsilon}^{2}.
\label{eq:dam}
\end{equation}
\vspace{-0.2mm}
Here $\varepsilon=.1$ is an insensitivity tolerance in percentile space. No reference/source amplitude is matched directly.

\subsection{Annealed Discrete Solver}

Eqs.~\eqref{eq:rel} and \eqref{eq:dam} induce a nonlocal, loopy interaction graph rather than a fixed-order Markov chain, so standard Viterbi recursion does not apply. We optimize
\begin{equation}
\mathcal L=\mathcal L_{\rm rel}+\lambda_{\rm dam}\mathcal L_{\rm dam}+\mathcal L_{\rm tr},
\label{eq:solver}
\end{equation}
where $\mathcal L_{\rm tr}$ denotes the usual concatenative-synthesis transition regularization, including MFCC/delta-MFCC boundary compatibility and simple continuation, repetition, and reuse preferences. We use $\lambda_{\rm dam} = 1$. The $\mathrm{CRG}_{\mathrm{nodam}}$, $\mathrm{FO}_{\mathrm{nodam}}$, and $\mathrm{VMO}_{\mathrm{nodam}}$ ablations in the experiments section set $\lambda_{\rm dam}=0$.

We use eight parallel restarts and twelve random-order coordinate sweeps, adding Gumbel perturbations whose temperature decreases linearly from $0.08$ to zero, and retain the lowest-cost solution. This nonlocal discrete search does not guarantee a global optimum.

\vspace{-2mm}
\section{Experiments}

\begin{table*}[!t]

\centering
\captionof{table}{Stable Audio~3 results on 248 evaluation triplets. Pair-level metrics first average the three neural seeds of each $(S,R,T)$ triplet; $KL_T$, $IS$, and $FD_{V/C/P}$ are set-level. \textbf{Bold} marks the best non-shortcut source-carrier result. \underline{Underlining} marks a numerical winner that is principally a shortcut or control bound. $^\dagger$ marks persistence of a directly exposed waveform. Arrows indicate the favorable direction for each evaluation role; Leakage metrics are diagnostic rather than minimization objectives, so numerical winners are not highlighted.}
\label{tab:main}

\renewcommand{\arraystretch}{0.90}
\setlength{\tabcolsep}{0.55pt}
\vspace{1.5mm}

\begin{tabular}{lrrrrrrrrrrrrrr}
\toprule
& \multicolumn{4}{c}{Temporal}
& \multicolumn{3}{c}{Acoustic}
& \multicolumn{4}{c}{Leakage}
& \multicolumn{1}{c}{Prompt}
& \multicolumn{2}{c}{Distribution} \\
Condition &
$L_T\downarrow$ & $L_R\downarrow$ & $O_T\uparrow$ & $O_R\uparrow$ &
$M_T\uparrow$ & $C_T\uparrow$ & $KL_T\downarrow$ &
$C_S\downarrow$ & $R_S^{2s}\downarrow$ & $C_R\downarrow$ & $R_R^{2s}\downarrow$ &
$C_P\uparrow$ & $IS\uparrow$ & $FD_{V/C/P}\downarrow$ \\
\midrule

None
& .086 & .094
& .354 & .305
& .913 & .540 & 1.806
& .542 & .167 & .177 & .118
& \underline{.246} & 8.83
& 8.82/.413/52.04
\\

\midrule
\multicolumn{15}{l}{\textit{Direct reference-audio controls}}\\

Ref $R$
& .083 & \underline{.059}
& \underline{.402} & \underline{.421}
& \underline{.939} & .524 & 2.196
& .511 & .170 & .243 & .365\rlap{$^\dagger$}
& .227 & 7.97
& 11.24/.405/56.74
\\

Mix $S+R$
& \underline{.080} & .084
& .395 & .347
& .927 & \underline{.562} & \underline{1.727}
& .590 & .303 & .198 & .196\rlap{$^\dagger$}
& .240 & 9.21
& 7.49/\underline{.354}/47.31
\\

\midrule
\multicolumn{15}{l}{\emph{Source-carrier conditions -- central comparison}}\\

Raw $S$
& .088 & .095
& .380 & .322
& .923 & .552 & 1.865
& .596 & .353\rlap{$^\dagger$} & .190 & .137
& .234 & \underline{9.34}
& 7.53/.354/47.34
\\

Shuffle
& .091 & .098
& .378 & .319
& .921 & .547 & \textbf{1.733}
& .588 & .299 & .196 & .137
& .224 & 8.49
& 6.70/.371/46.65
\\

Pointwise-SR
& .095 & .101
& .389 & .335
& .921 & .523 & 2.011
& .552 & .286 & .191 & .177
& .214 & 8.47
& 6.79/.372/47.22
\\

CRG (ours)
& .096 & .097
& \textbf{.394} & .343
& .921 & .515 & 2.110
& .545 & .307 & .187 & .151
& .225 & 8.63
& \textbf{6.40}/.389/47.35
\\

FO (ours)
& \textbf{.087}
& \textbf{.086}
& .379
& \textbf{.343}
& .925
& \textbf{.551}
& 1.737
& .592 & .299 & .195 & .144
& \textbf{.237}
& 8.70
& 7.29/\textbf{.365}/\textbf{45.66}
\\

VMO (ours)
& .087 & .087
& .389 & .341
& \textbf{.926}
& .543
& 1.764
& .584 & .300 & .194 & .145
& .233
& \textbf{8.80}
& 7.10/.368/46.54
\\

\bottomrule
\end{tabular}

\vspace{-1.5mm}
\captionof{table}{DAM ablation on the same 248 triplets. Each $\mathrm{nodam}$ condition removes Eq.~\eqref{eq:dam}. All other recurrence structure is unchanged. \textbf{Bold} marks the favorable value within each matched pair. $FD_{V/C/P}$ denotes VGGish/CLAP/PANN Fr\'echet distance.}
\label{tab:dam}
\setlength{\tabcolsep}{2.2pt}
\vspace{1.5mm}
\begin{tabular}{lrrrrrrrrr}
\toprule
& \multicolumn{2}{c}{Temporal}
& \multicolumn{1}{c}{Acoustic}
& \multicolumn{2}{c}{Leakage}
& \multicolumn{1}{c}{Prompt}
& \multicolumn{3}{c}{Distribution} \\
Condition &
$L_R\downarrow$ & $O_R\uparrow$ &
$C_T\uparrow$ &
$R_S^{2s}\downarrow$ & $R_R^{2s}\downarrow$ &
$C_P\uparrow$ &
$KL_T\downarrow$ & $IS\uparrow$ &
$FD_{V/C/P}\downarrow$ \\
\midrule
CRG$_{\mathrm{nodam}}$
& .110
& .328
& .491
& .310
& .141
& .216
& 2.197
& 8.23
& 6.62/.410/50.14
\\

CRG
& \textbf{.097}
& \textbf{.343}
& \textbf{.515}
& .307
& .151
& \textbf{.225}
& \textbf{2.110}
& \textbf{8.63}
& \textbf{6.40}/\textbf{.389}/\textbf{47.35}
\\

\midrule

FO$_{\mathrm{nodam}}$
& .096
& .320
& .548
& .306
& .142
& .233
& 1.821
& \textbf{8.82}
& \textbf{7.18}/\textbf{.362}/46.59
\\

FO
& \textbf{.086}
& \textbf{.343}
& \textbf{.551}
& .299
& .144
& \textbf{.237}
& \textbf{1.737}
& 8.70
& 7.29/.365/\textbf{45.66}
\\

\midrule

VMO$_{\mathrm{nodam}}$
& .096
& .324
& \textbf{.546}
& .294
& .139
& .230
& 1.770
& \textbf{9.08}
& \textbf{6.79}/\textbf{.359}/\textbf{46.49}
\\

VMO
& \textbf{.087}
& \textbf{.341}
& .543
& .300
& .145
& \textbf{.233}
& \textbf{1.764}
& 8.80
& 7.10/.368/46.54
\\
\bottomrule
\end{tabular}
\end{table*}

\vspace{-1.5mm}
\subsection{Setup}

We use ESC-50 \cite{piczak2015esc} and ESC-50-Voice \cite{okamoto2024vocal}. ESC-50-Voice contains vocal imitations, which we use as references $R$, for 31 classes of environmental sounds in ESC-50, which we use as targets $T$. The targets are used only for evaluation; for each $T$, we pair it with one corresponding vocal imitation $R$.

ESC-50 divides each class into five folds. Two Fold-1 targets per class are used to tune conditioning strength. All eight Fold-2 examples per class are then used as evaluation targets, while the source $S$ is sampled independently, without replacement, from the same class in Folds 3-5. Same-class source sampling serves as oracle retrieval, isolating synthesis from retrieval quality. The final evaluation therefore contains $31\times8=248$ distinct $(S,R,T)$ triplets. All inputs are 5-s mono audio at 22.05~kHz. Let $P$ denote the prompt \texttt{A realistic <class> environmental sound. TrackType: SFX}, where \texttt{<class>} is replaced at runtime by the class label of $S$.

We instantiate $G_\theta$ as the pretrained Stable Audio~3 Small-SFX \cite{evans2026stableaudio3}, using eight generation steps and three generation seeds per triplet. On Fold-1, we sweep init-noise levels $\{0.25,0.50,0.75\}$ and select $0.50$ for the final Fold-2 evaluation. No task-specific training or additional conditioning network is introduced. The RAG conditions are: \emph{None}, using text prompt only; \emph{Ref}, directly using $R$; \emph{Mix}, using equal-RMS $S+R$; \emph{Raw}, directly using $S$; \emph{Shuffle}, using block-shuffled $S$; \emph{Pointwise-SR}, using MFCC-nearest-grain-based pointwise concatenative synthesis; and \emph{CRG/FO/VMO}, using relational synthesis. Unless marked with the $_{\mathrm{nodam}}$ subscript, DAM is enabled for all relational-synthesis conditions.

\vspace{-1.5mm}
\subsection{Metrics}

Let $Y$ denote the final neural output of the pipeline. Recall that we started with $\mathcal T(Y)\to\mathcal T(R)\approx\mathcal T(T)$, the \textbf{temporal} constraints; $\mathcal A(Y)\to\mathcal A(T)$, the \textbf{acoustic} constraints; and $\Pi(Y,S),\Pi(Y,R)$, the \textbf{leakage/copying} constraints. For the acoustic constraint, we evaluate against $T$ rather than $S$: by setup, $T$ has the same class as $S$, while chasing $\mathcal A(S)$ may reward persistent copying. Since the final stage is also a text-to-audio generator, we additionally report standard prompt-alignment and distributional-quality checks.

\renewcommand{\arraystretch}{0.90}
\setlength{\tabcolsep}{2pt}
\noindent
\begin{tabularx}{\linewidth}{@{}p{0.33\linewidth} X@{}}
\toprule
\textbf{Metric} & \textbf{Computation} \\
\midrule
$\mathcal T$: $L_T,L_R\downarrow$
& Smoothed RMS-envelope L1 \cite{chung2024tfoley} \\
$\mathcal T$: $O_T,O_R\uparrow$
& Onset F1 (50 ms) \\
$\mathcal A$: $M_T\uparrow$
& Log-mel cosine \cite{bralios2024replication} \\
$\mathcal A$: $C_T\uparrow$
& CLAP audio cosine \cite{wu2023clap} \\
$\mathcal A$: $KL_T\downarrow$
& PANN KL \cite{kong2020panns} \\
$\Pi$: $R_S^{2s},R_R^{2s}$
& Best-match 2-s log-mel cosine \\
$\Pi$: $C_S,C_R$
& CLAP audio cosine \\
Prompt: $C_P\uparrow$
& CLAP audio--text cosine \\
Distribution: $IS\uparrow$
& PANN Inception Score \cite{kong2020panns} \\
Distribution: $FD_{V/C/P}\downarrow$
& VGGish/CLAP/PANN Fr\'echet distance \cite{kilgour2019fad} \\
\bottomrule
\end{tabularx}
\vspace{-1mm}

Leakage similarities are diagnostic rather than monotone objectives: high similarity is undesirable, but minimizing similarity itself is not the goal. Pair-level metrics first average the three neural seeds for each $(S,R,T)$ triplet before aggregation.

\vspace{-1mm}
\subsection{Results and Discussion}
We will read the results in the same order as the evaluation logic.

\noindent\textbf{1. Temporal transfer $\mathcal T(Y)\to\mathcal T(R)\approx\mathcal T(T)$:}
The central question is whether use of relational information from $R$ can improve temporal organization when the generator still receives acoustic material only from $S$. Relative to the source-carrier controls, all three relational constructions increase reference-onset agreement: $O_R=.322/.319/.335$ for Raw/Shuffle/Pointwise-SR, compared with $.343/.343/.341$ for CRG/FO/VMO. FO gives the clearest complementary Event-L1 result: relative to Raw, $L_R$ decreases from $.095$ to $.086$ while $O_R$ rises from $.322$ to $.343$. Both changes are significant (paired 95\% CIs, FO$-$Raw:
$\Delta L_R\in[-.012,-.007]$, $\Delta O_R\in[.008,.034]$). The direct-reference controls confirm that still stronger timing can be obtained when $R$ itself is exposed to the generator ($L_R=.059,O_R=.421$), but those conditions have access to different acoustic information and serve as reference-access controls rather than the central source-carrier comparison.

\noindent\textbf{2. Acoustic-fidelity $\mathcal A(Y)\to\mathcal A(T)$:}
The useful temporal gain should preserve target-class acoustic fidelity carried by $S$ and requested by $P$. FO is the most stable relational realization on this axis: Raw$\rightarrow$FO essentially preserves target CLAP ($.552\rightarrow.551$), slightly raises target log-mel similarity ($.923\rightarrow.925$), preserves prompt CLAP ($.234\rightarrow.237$), and improves PANN KL ($1.865\rightarrow1.737$). Thus the timing gain does not require moving the generated event identity toward the vocal reference.

\noindent\textbf{3. Persistent Leakage $\Pi(Y,S),\Pi(Y,R)$:}
The strongest evidence for the structural bottleneck appears when timing and persistence are read together. FO reduces local source persistence relative to Raw, $R_S^{2s}=.353\rightarrow.299$ (paired 95\% CI $[-.068,-.039]$), while its reference persistence remains near source-carrier controls: $C_R=.195$ and $R_R^{2s}=.144$ for FO, compared with $.190/.137$ for Raw and $.196/.137$ for Shuffle. Direct Ref, by contrast, undesirably reaches $R_R^{2s}=.365$. Hence improved reference timing does not require passing the reference waveform through the generator, while the source is also not simply replayed locally.

\noindent\textbf{4. Prompt and distributional bounds.}
FO preserves the pretrained generator's semantic operating point within the source-carrier comparison: prompt CLAP is essentially unchanged from Raw ($.234\rightarrow.237$). Its VGGish FAD improves from $7.53$ to $7.29$, PANN Fr\'echet distance from $47.34$ to $45.66$, and PANN KL from $1.865$ to $1.737$; CLAP FAD changes moderately from $.354$ to $.365$, and IS from $9.34$ to $8.70$. These set-level quantities are therefore treated as bounds rather than the primary optimization target: the desired temporal, acoustic, and leakage behavior is not purchased by a large collapse in semantic or distributional behavior.

\noindent\textbf{5. Directed-amplitude-motion mechanism ablation.}
The DAM ablation is deliberately internal to each relational structure. Enabling DAM changes $L_R$ from $.110\rightarrow.097$, $.096\rightarrow.086$, and $.096\rightarrow.087$ for CRG, FO, and VMO, while $O_R$ changes from $.328\rightarrow.343$, $.320\rightarrow.343$, and $.324\rightarrow.341$. All six temporal deltas have paired 95\% CIs excluding zero. Acoustic and prompt alignment are generally preserved or improved across the matched pairs, whereas leakage changes are comparatively small and the set-level effects are mixed. This is consistent with the intended reduction of the acoustic-temporal trade-off: DAM strengthens temporal transfer without forcing cross-domain envelope values to coincide, while the acoustic fidelity measures improve rather than being traded away. Its benefit is therefore complementary to recurrence structure.

\end{document}